 \documentclass[aps,pre,superscriptaddress, nofootinbib, longbibliography, notitlepage, showkeys]{revtex4-1}
\makeatletter
\renewcommand\frontmatter@abstractwidth{\dimexpr\textwidth-1in\relax}
\makeatother
\usepackage{braket}

\usepackage[bitstream-charter, greekfamily=default]{mathdesign}

\usepackage[utf8]{inputenc} 	

\usepackage[margin=1.5 in]{geometry} 

\usepackage[english]{babel}		
\usepackage{amsmath,amssymb,amsthm}	
\usepackage{mathtools}
\usepackage{graphicx}			
\usepackage{hyperref}			
\usepackage{color}			
\usepackage{ds font}
\usepackage{array}
\usepackage{booktabs}
\usepackage{tabularx}
\usepackage{makecell}
\usepackage{siunitx} 
\usepackage{multirow}

\newcommand{\inlinemaketitle}{{\let\newpage\relax\maketitle}}

\begin{document}

\title{The Case for Long-Only Agnostic Allocation Portfolios}

\author{{Pierre-Alain Reigneron, Vincent Nguyen, Stefano Ciliberti,} \\
{Philip Seager, Jean-Philippe Bouchaud}\\
{Capital Fund Management} \\
{23 rue de l’Université, 75007 Paris, France}
}



\begin{abstract}
We advocate the use of Agnostic Allocation for the construction of long-only portfolios of stocks. We show that Agnostic Allocation Portfolios (AAPs) are a special member of a family of risk-based portfolios that are able to mitigate certain extreme features (excess concentration, high turnover, strong exposure to low-risk factors) of classical portfolio construction methods, while achieving similar performance. AAPs thus represent a very attractive alternative risk-based portfolio construction framework that can be implemented in different situations, with or without an active trading signal.
\end{abstract}
\let\newpage\relax%
\maketitle

\section{Introduction}

Risk-Based Portfolios (RBPs) rely on a forecast-agnostic approach to investing, and they have risen in popularity since the global financial crisis. Their success reflects a growing disbelief in active managers' ability to deliver alpha, together with an increased emphasis on risk as a core component of investment policies. These RBPs all seek to efficiently capture some excess premium in one or multiple asset class(es) by factoring in risk-related quantities, without any specific views on expected returns. 

The present paper addresses the case of long-only portfolios of stocks. We suggest that many well documented RBPs belong to a ``risk-based continuum", from which we isolate the Agnostic Allocation Portfolio (AAP) \cite{ARP} as a special case of particular interest. Using an eigenvector decomposition, we motivate a natural extension of AAP that we call ``Eigen-Sparse'', that avoids investing on low risk, high-cost modes that should only be traded when specific, high conviction forecasts are available.   

We show from the point of view of Sharpe ratio and total return that AAPs can compete with or even outperform standard RBPs, while significantly mitigating documented flaws of methods relying on covariance matrices, namely over-concentration and excess turnover, as well as exposure to low-risk factors.

\section{Risk-Based Portfolios: a Short Primer}

Certain RBPs rely on explicit weighting schemes such as the market capitalization-weighted portfolio, or the equal risk budget portfolio. Others rely on the optimization of a risk-related objective function. In most cases, such optimization schemes do not have closed-form solutions when investment constraints (such as the long-only constraint or maximum position constraints) are added, see e.g. \cite{Michaud,MV2,CAL}.

In this section we briefly review some well-documented RBPs. We will consider a universe of $N$ stocks $i=1,\dots, N$, with price $S_i(t)$ at time $t$. The (daily) return of stock $i$, $r_i(t)$, is defined as:
\begin{equation}\label{ret}
r_i(t)=\frac{S_i(t)}{S_i(t-1)}-1.
\end{equation}
These returns are characterized by their covariance matrix $\mathbf{C}$, defined (theoretically) as
\begin{equation}\label{covar}
\begin{aligned}
&  C_{i,j}=\mathds{E}[(r_i-\mathds{E}[r_i])(r_j-\mathds{E}[r_j])].
\end{aligned}
\end{equation}
The square-root of the diagonal elements define the vector of stock volatilities: ${\bf \sigma}_i = \sqrt{C_{i,i}}$. Of course, the `true' (forward-looking) covariance matrix $\mathbf{C}$ is not known at the time of investment and has to be ``guesstimated'' as best as possible from past data -- see below in Appendix B for more on this topic. 

We will call $w_i$ the weight of stock $i$ in the portfolio. This weight is considered under fully-invested, long-only constructions, such that $0 \leq w_i \leq 1$ and $\sum_{i=1}^N w_i=1$. 
 
\subsection{Explicit weightings}

\begin{itemize}
\item Market cap --- The market-cap ``MC" index is the most trivial of all as it simply allocates stock weights according to their market cap $\it{M}_i$:
\begin{equation}\label{MC}
{w}_i=\frac{\it{M}_i}{\sum_j\it{M}_j}
\end{equation}
The market-cap index is known to be mean–variance efficient if the CAPM assumptions are valid. These are however well-known to be very far from realistic (see for example \cite{Fernandez} and refs. therein)

\item Equal weight ---
The equal weight ``1/N" portfolio \cite{1N} allocates uniformly among all stocks:
\begin{equation}\label{1overN}
{w}_i=\frac{1}{N}
\end{equation}
The portfolio is maximally diversified in terms of instrument weights. More precisely, it minimizes the Herfindal index $H$, defined as
\begin{equation}
H := \sum_{i=1}^N {w}_i^2 =\frac{1}{N}
\end{equation}
It is the efficient portfolio assuming that stocks all have the same expected return and volatility, and that all pairwise correlations are equal.

\item Equal volatility ---
Another simple portfolio construction amounts to allocating to a stock a weight that is inversely proportionally to its volatility:
\begin{equation}\label{EV}
{w}_i=\frac{\sigma_i^{-1}}{\sum_j \sigma_j^{-1}} 
\end{equation}
It is the efficient portfolio assuming that stocks all have the same Sharpe ratio, and that all pairwise correlations are zero.
\end{itemize}

\subsection{Objective Function Based Portfolios}

These methods often minimize quadratic forms involving the covariance matrix of stocks returns.

\begin{itemize}
\item Minimum Variance ---
The Minimum Variance portfolio (MVP) (\cite{MV0,MV1}) is the portfolio minimizing the ex-ante volatility, taking into account the correlations between stocks. It is thus an objective function-based portfolio, and follows from the following optimization program: 
\begin{equation}\label{MVeq}
\begin{aligned}
{\bf w}^* = & \enskip \operatorname*{arg\,min}_{\bf w} & & {\bf w}^\top \mathbf{C}{\bf w} \\
& \text{subject to} & & \mathds{1}^\top {\bf w}=1,\\
&&& 0 \leq  w_i \leq 1
\end{aligned}
\end{equation}
where $\mathds{1}$ is the vector $(1,1,..,1)$ of size $N$, $\mathbf{C}$ is the covariance matrix for stocks returns. This portfolio is efficient if expected returns are equal for all stocks. One well-documented feature \cite{MV2,CAL} feature of this portfolio is its propensity to generate a severe concentration of weights on a small subset of stocks -- much smaller than the available trading universe. This is a characteristic shared also by other portfolios relying on the minimization of a quadratic form involving the covariance matrix.

\item Maximum Diversification ---
The Maximum Diversification portfolio (MDP) introduced in \cite{MDP}, is the portfolio maximizing the diversification ratio:
\begin{equation}\label{MDPeq}
\begin{aligned}
{\bf w}^* = & \enskip \operatorname*{arg\,max}_{\bf w} & &  \frac{{\bf w} \cdot {\bf \sigma}}{\sqrt{{\bf w}^\top {\bf C} {\bf w}}} \\
& \text{subject to} & & \mathds{1}^\top {\bf w}=1,\\
&&& 0 \leq  w_i \leq 1
\end{aligned}
\end{equation}
Intuitively, the diversification ratio compares the risk of a portfolio assuming that stocks are uncorrelated (namely, ${\bf w} \cdot {\bf \sigma}$) with the actual risk of the same portfolio, but accounting for correlations. As this ratio becomes larger, the stocks that compose this portfolio become more ``effectively decorrelated''.

The MDP can also be seen as the efficient portfolio when all stocks share the same expected Sharpe ratio, or said differently, when the expected return of each stock is equal to some common coefficient times its volatility.

\item Equal Risk Contribution ---
Yet another construction was proposed in \cite{ERC} as the Equal Risk Contribution portfolio (ERC). Using Euler's theorem on homogeneous functions (here, the portfolio variance as a function of the weights) the final allocation problem reads
\begin{equation}\label{ERCeq}
\begin{aligned}
{\bf w}^* \enskip &  \text{such that} & &  w_i \left({\bf C} {\bf w}\right)_i = w_j \left({\bf C} {\bf w}\right)_j, \enskip \forall i,j\\
& \text{subject to} & & \mathds{1}^\top {\bf w}=1, \\
&&& 0 \leq  w_i \leq 1
\end{aligned}
\end{equation}

\end{itemize}

\section{A Target Portfolio Approach}

\subsection{General Setting} 

The optimization of objective functions, like the ones considered in the previous section, must respect some constraints such as the positivity of the weights (long-only). Such optimization problems lack  closed-form solutions, but they can be reformulated as a more general tracking-error problem. One must first construct the optimal portfolio in the absence of constraints. We call this the ``target portfolio'' ${\bf w}^{\text{t}}$. The optimal long-only portfolio  ${\bf w}^*$ can then be described as the long-only portfolio that minimizes its tracking-error to  ${\bf w}^{\text{t}}$:  
\begin{equation}\label{optrb}
\begin{aligned}
{\bf w}^*= & \enskip \operatorname*{arg\,min}_w  & &  ({\bf w}-{\bf w}^{\text{t}})^\top \mathbf{C} ({\bf w}-{\bf w}^{\text{t}})\\
& \text{subject to} & & \enskip 0 \leq  w_i \leq 1
\end{aligned}
\end{equation}
For example, in the case of the MVP, the target portfolio is readily obtained as
\begin{equation}\label{MVuncsol}
\begin{aligned}
{\bf w}^{\text{t}}_{\text{MV}}= \omega \mathbf{C}^{-1} \mathds{1}
\end{aligned}
\end{equation}
Since the non-constrained target portfolio $w^{\text{t}}_{MV}$ is a long-short portfolio, the mean-variance optimal portfolio (in the case where $\mathds{1}$ is used as an agnostic forecast) tends to allocate heavily to ``market neutral" configurations. $\omega$ is a scaling factor, which we set such that the unconstrained solution has a net exposure of $100\%$ in order to give the problem a dimension. The optimal portfolio ${\bf w}^*$ is of arbitrary size and thus needs to be re-scaled to the trading level of the portfolio. 
\begin{equation}\label{alphascale}
\begin{aligned}
\omega=\frac{1}{\sum_{ij}C^{-1}_{ij}}=\frac{1}{\mathds{1}^\top \cdot \mathbf{C}^{-1}\cdot \mathds{1}} 
\end{aligned}
\end{equation}
The MDP problem corresponds to the following target portfolio:
\begin{equation}\label{MDPunconstrained}
\begin{aligned}
w^{\text{t}}_{\text{MDP}}= \omega \mathbf{C}^{-1} \bf{\Sigma} \mathds{1},
\end{aligned}
\end{equation}
where $\bf \Sigma$ is the $N$ by $N$ diagonal matrix such that $\Sigma_{i,i}=\sigma_{i}$. $w^{\text{t}}_{\text{MDP}}$ is thus an unconstrained solution to the mean-variance optimization problem applied to a vector of implicit forecast returns ${\bf \Sigma} \mathds{1}$ corresponding to the stocks' volatilities. 

\subsection{A Continuum of Target Portfolios}

Reformulating risk-based indexing methods as the minimization of a tracking error with respect to a unconstrained target portfolio enables us to generate a whole family of potentially interesting long-only portfolios. This idea can be generalized to the MVP or MDP by writing the expression for the target portfolio as
\begin{equation}\label{continuum}
\begin{aligned}
{\bf w}^{\text{t}}_{a,b,c}= \omega \mathbf{C}^{-a} {\bf \Sigma}^{b} {\bf M}^c \mathds{1},
\end{aligned}
\end{equation}
where $\bf M$ is the $N$ by $N$ diagonal matrix such that ${\bf M}_{i,i}=M_i$ is the market capitalization of stock $i$. The corresponding long-only portfolio, ${\bf w}^*_{a,b,c}$ is obtained from the optimization problem (\ref{optrb}) with ${\bf w}^{\text{t}}_{a,b,c}$ as a target. The code for this algorithm and the documentation will be made available by the authors upon request. 

Most of the previous risk based portfolios can be recovered in this general setting. For example, $a=b=0$ and $c=1$ corresponds to the market cap index portfolio;  $a=b=c=0$ is the equal weight portfolio; $a=c=0$, $b=-1$ is the equal volatility portfolio. $b=c=0$, $a=1$ is the standard mean-variance portfolio whereas
$c=0$, $a=b=1$ corresponds to the maximum diversification portfolio. More generally, the 3 parameters $a,b,c$ have the following interpretations: 
\begin{itemize}
\item $a$: values close to zero make the portfolio blind to covariance. Values close to 1 will allocate to low risk -- typically market-neutral -- combinations of stocks (in the diagonal basis, the inverse covariance matrix acts as a multiplication by the inverse of the variance of the corresponding mode). This creates target portfolios for which the long-only constraint is more acute, which in turn translates into a more concentrated final allocation. 

\item $b$ is a volatility affinity parameter: increasing $b$ is tantamount to believing that higher volatility stocks have larger expected returns. 

\item $c$ is a market cap affinity parameter. Values $\approx 1$  indicate a preference for larger market capitalization. This parameter can be used in practice to mitigate transaction costs by favouring large capitalisations. 
\end{itemize}

\subsection{Risk decomposition}\label{risk_decomp}

It is useful to recall the standard decomposition of risk onto the eigenmodes of the covariance matrix $\bf{C}$. Introducing the eigenvalues $\lambda_k$ ($k=1,\dots,N$), ranked in decreasing order, and $\bf{u}_k$ as the corresponding eigenvectors, one has:
\begin{equation}\label{eigen}
\begin{aligned}
\bf{C} = \sum_k \lambda_k \mathbf{u}_k^\top \mathbf{u}_k
\end{aligned}
\end{equation}
Introducing an effective predictor ${\bf p}$ corresponding to Eq. (\ref{continuum}), defined as $p_i := \sigma_i^b M_i^c$, the weights $w_i$ can be written as
\begin{equation}\label{eigen2}
\begin{aligned}
w_i = \omega \sum_k \lambda_k^{-a} (\mathbf{u}_k \cdot \mathbf{p}) (\mathbf{u}_k)_i,
\end{aligned}
\end{equation}
so that the risk $R^2$ of the target portfolio is given by
\begin{equation}\label{eigen3}
\begin{aligned}
R^2 = \omega^2 \sum_k \lambda_k^{1-2a} (\mathbf{u}_k \cdot \mathbf{p})^2.
\end{aligned}
\end{equation}
This last equation lends itself to an insightful interpretation that will be discussed further in the next section. One sees that for $a < 1/2$, the contribution of the $k$th mode to the risk is given by the projection of the predictor on that mode times a factor that increases with the risk $\lambda_k$ of that mode. Conversely, for $a > 1/2$, the natural projection of the predictor is enhanced as $\lambda_k$ is decreased. This is the case, for example, of the MVP for which $a=1$. It is well known that Markowitz \footnote{In this article we use "mean-variance" and "Markowitz" interchangeably.} tends to overallocate to small risk modes. Finally, for $a=1/2$, the risk of the portfolio is allocated proportionally to the projection of the predictor, with no further bias towards large risk or small risk modes. 

\subsection{Statistical Exploration of the Risk-Based Continuum}

In this subsection, we study the influence of the parameter $a$ on characteristics that are of paramount importance in practice: beta/correlation to the benchmark, portfolio concentration and turnover of the long-only risk based portfolios, and the appetite for short positions in the corresponding target portfolios. 

For the purpose of this discussion, we work with a pool of $\sim 2000$ US stocks with an uninterrupted price history over 3500 days (Aug 2005- Dec 2018). This introduces a survivorship bias in our pool of instruments; however, this is not an issue for the study of the statistical properties of risk-based portfolios. In addition, this setup is not polluted by the entries or exits in the trading pool that affect actual portfolios (but which are duly included in section 4 on the empirical results). 

In order to show de-noised, readable results, we use a bootstrapping method: we select $N_{\text{boot}}=10$ random samples of 250 stocks in the available universe. For each of these samples, we compute, for all rebalancing days $t$ (here, every other month end) the target portfolio corresponding to a given value of $a \in [0,1]$ where $b=c=0$. The covariance matrix $\bf C$ is the empirical covariance matrix measured over the previous period $[t-2-T,t-2]$. A shift of two days ensures causality, and $T$ is chosen as $2N=500$. We run two tests: a)  one using the raw covariance matrix and b) the second using a ``cleaned'' covariance matrix, based on the cross-validated RIE method recalled in Appendix B. 

\subsubsection{Volatility, Correlation and Beta}

\begin{figure}
\begin{center}
\includegraphics[width=13cm]{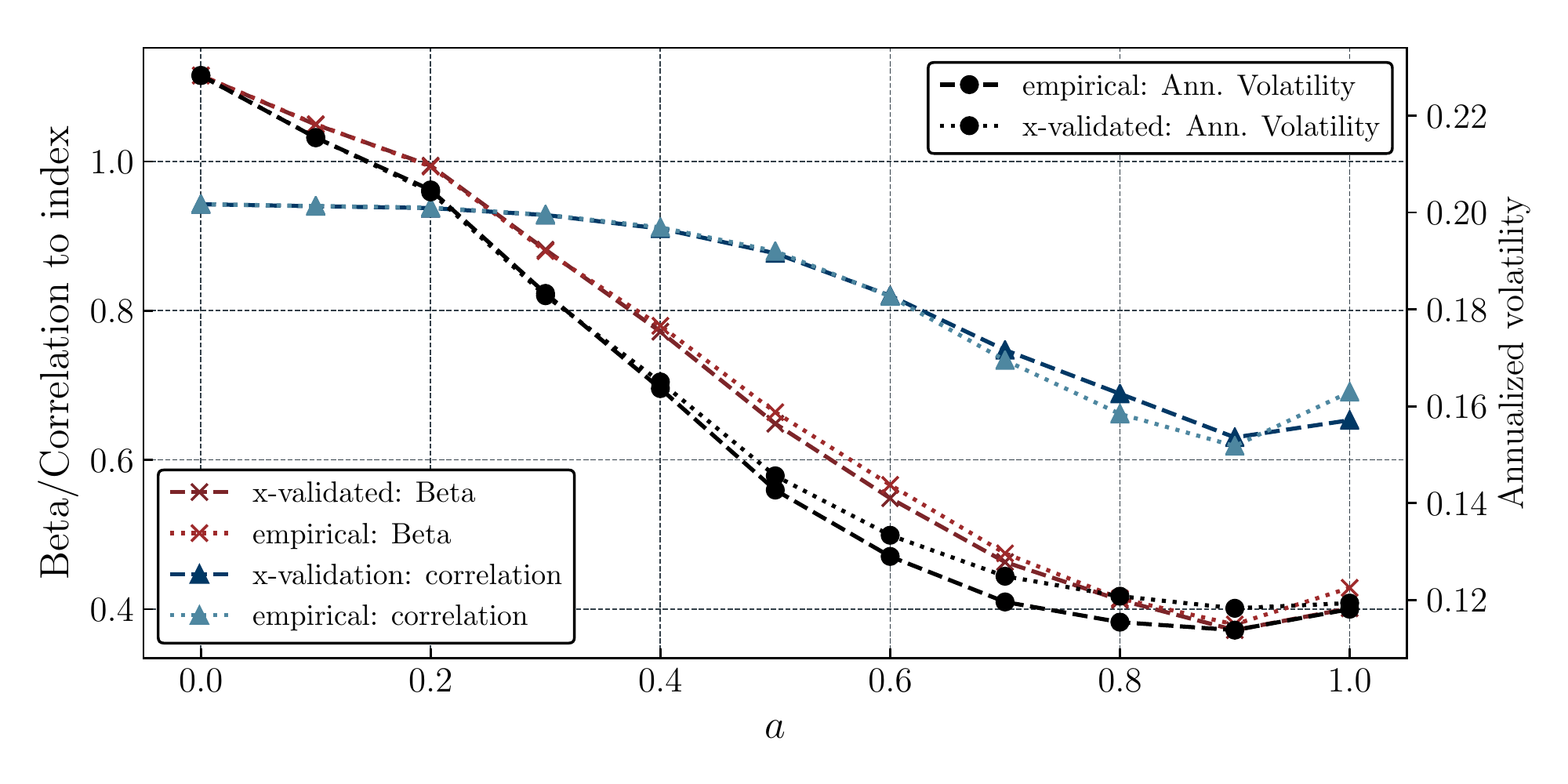}
\caption{Annualized daily volatility, beta, and correlation to the market cap benchmark as a function of $a$, for $b=c=0$.}
\label{beta}
\end{center}
\end{figure}

An expected characteristic of Risk-Based Portfolios is a reduction in volatility as $a$ tends to unity. This is confirmed by the results in  Figure \ref{beta} (black symbols). Correlation to the market is a monotonic decreasing function of $a$, which translates into a market beta behaving also as a monotonic decreasing function of $a$. 

\subsubsection{Short Positions of Target Portfolios}

\begin{figure}
\begin{center}
\includegraphics[width=10cm]{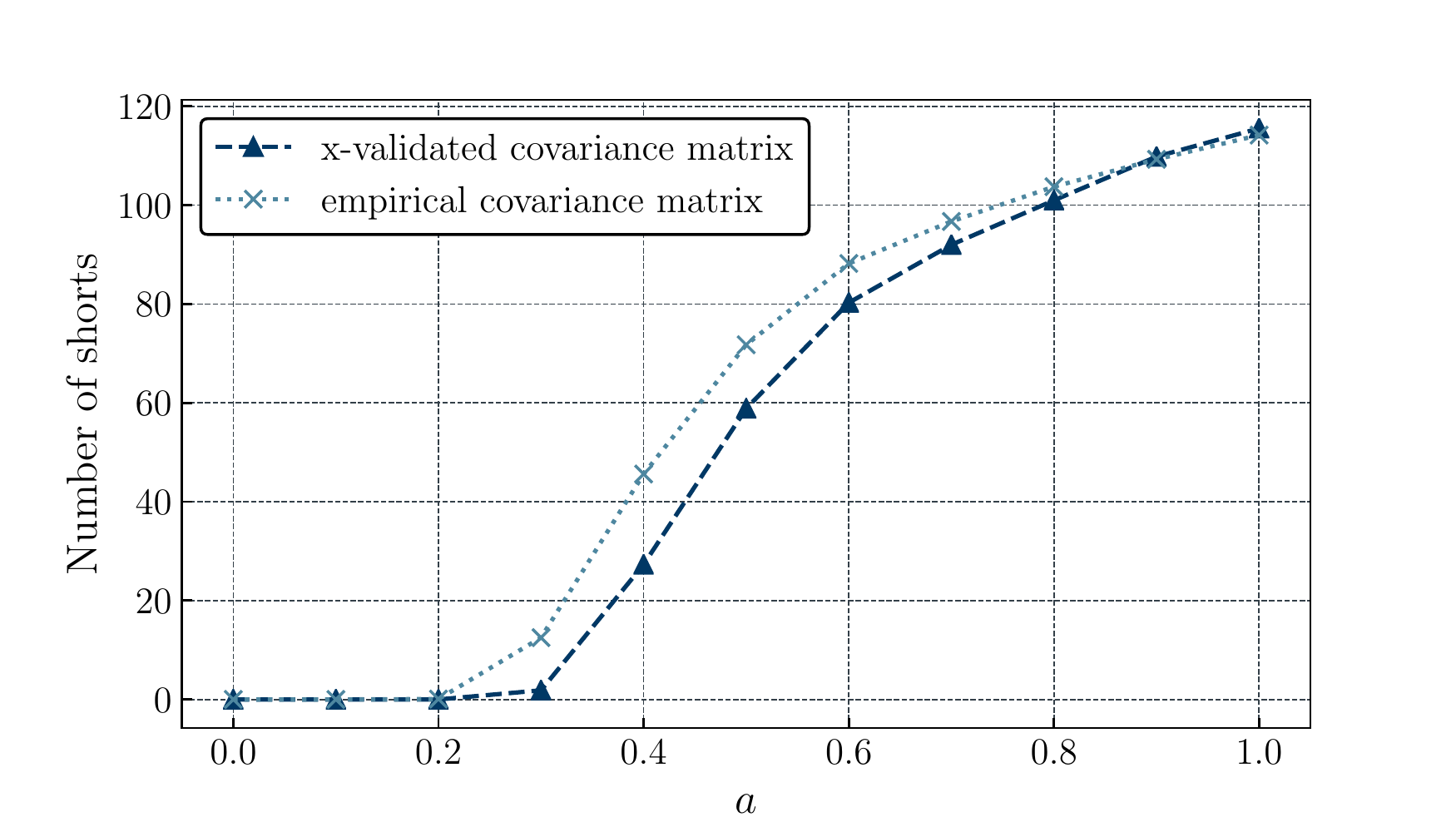}
\caption{Average number of short positions  for risk-based target portfolios, as a function of $a$, for $b=c=0$. This quantity is not very sensitive to the cleaning of the empirical covariance matrix.
}
\label{shorts}
\end{center}
\end{figure}

 Target portfolios build a larger and larger number of short positions as $a$ increases. This is shown in Figure \ref{shorts}: for $a \lesssim 0.2$, hardly any short position is taken, so the long-only portfolio will track almost exactly the target portfolio. As $a$ reaches the MVP value of $1$, the number of short positions in the target portfolio is almost $N/2$. 

\subsubsection{Concentration}

Concomitantly to the appetite for short positions of the target portfolio, the concentration of long-only optimized portfolios increases with $a$. Concentration of long-only Mean-Variance Portfolios is a documented feature \cite{Michaud,CAL} shared by the Maximum Diversification Portfolio, which is in itself an interesting paradox.

Here, we characterize the concentration of a portfolio by the so-called effective number of positions  $N_{\text{eff}}:= 1/H$where $H$ is the Herfindahl index $H = \sum_i w_i^2$. For an equally-weighted portfolio, $w_i=1/N$ and 
therefore $N_{\text{eff}}=N$.

\begin{figure}
\begin{center}
\includegraphics[width=10cm]{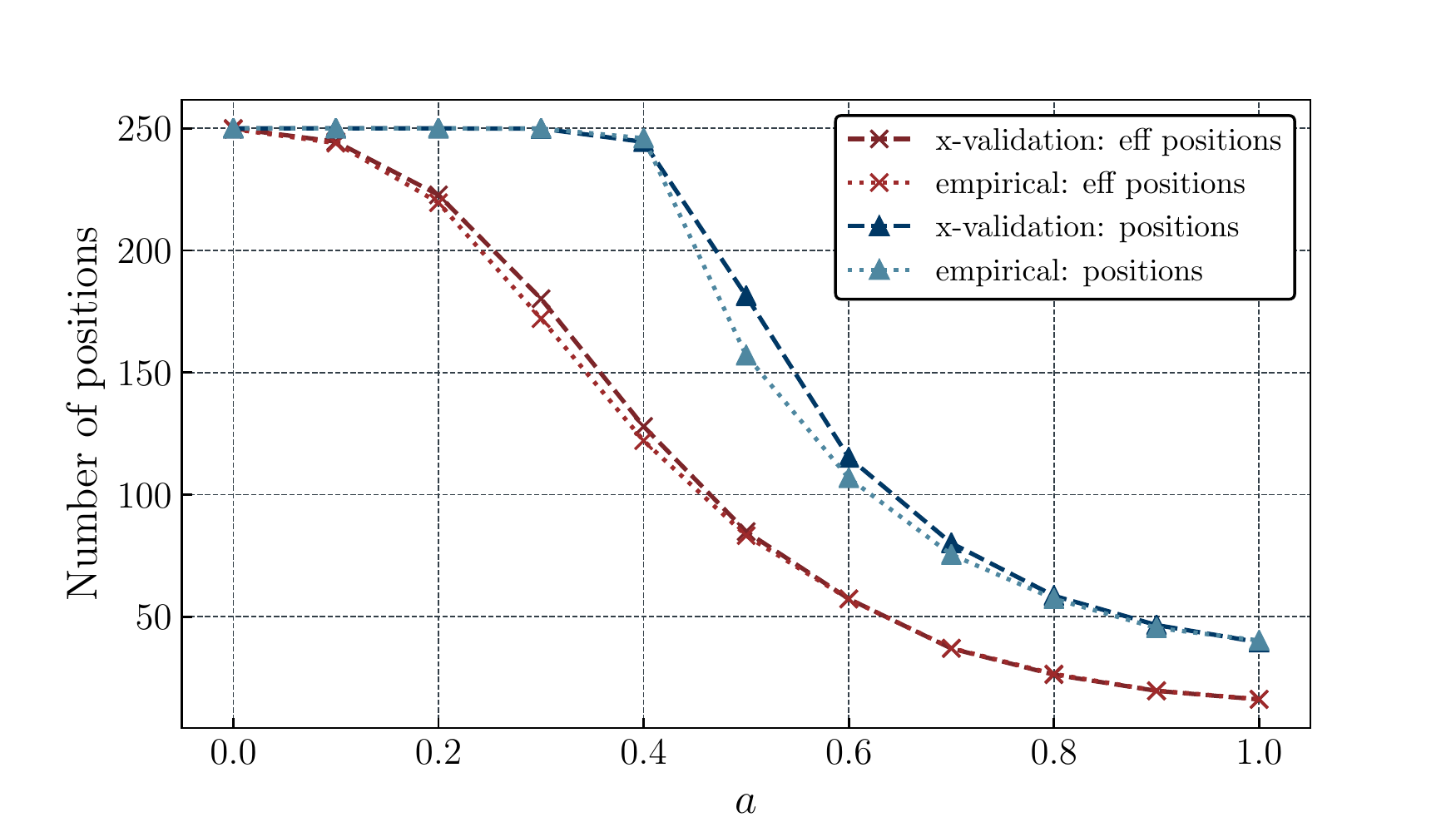}
\caption{Effective number of positions $N_{\text{eff}}$ and number of non-zero positions for risk-based optimized portfolios, as a function of $a$, for $b=c=0$. Again, these quantities are not very sensitive to the cleaning of the empirical covariance matrix.}
\label{sparsity}
\end{center}
\end{figure}

Figure \ref{sparsity} shows the effective number of positions $N_{\text{eff}}$ of a risk-based portfolio as a function of $a$. We see that it is a monotonic, decreasing function of $a$ that starts at $N_{\text{eff}}=N=250$ for $a=0$ (corresponding to equal weights) and ends at a rather low value $N_{\text{eff}} \approx 15 \sim N/20$ for $a=1$ (MVP). A similar dependence on $a$ is found for different values of $b$ and $c$. 

\subsubsection{Turnover}

\begin{figure}
\begin{center}
\includegraphics[width=10cm]{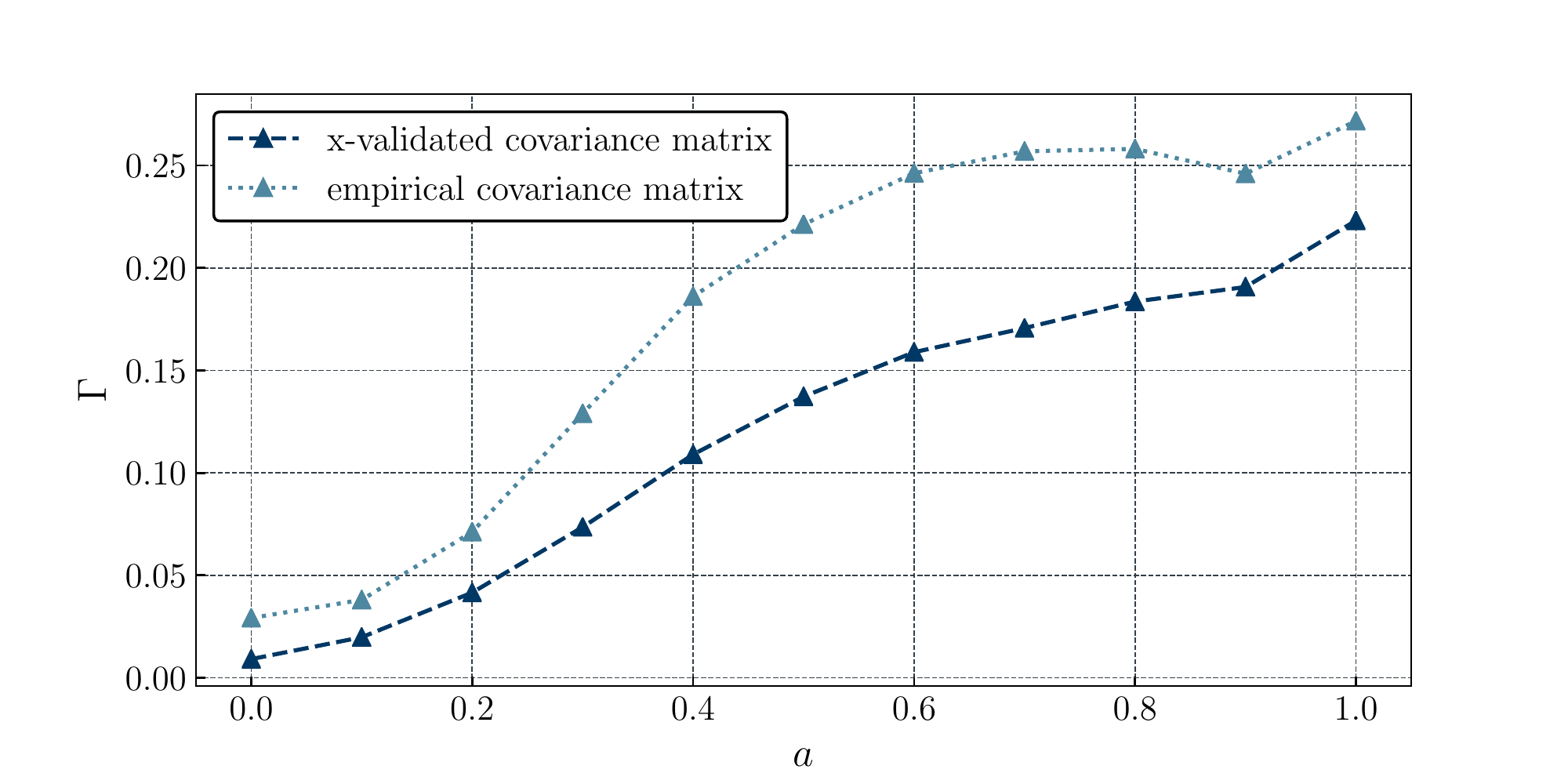}
\caption{Speed of change $\Gamma$ of the target portfolio as a function of $a$. As expected, $\Gamma$ is an increasing function of $a$. Note that using a raw covariance matrix leads to larger values of $\Gamma$ compared to the case of a RIE-cleaned covariance matrix.}
\label{trading}
\end{center}
\end{figure}

In addition to concentration, risk-based portfolios that rely on the inverse covariance matrix are known to lead to excessive turnover, see for example \cite{RBI}. Using the same bootstrapping method as above, we can investigate the influence of $a$ on turnover. We measure the speed $\Gamma$ at which the target portfolio changes as the distance between two consecutive portfolios, at times $T_{n-1}$ and $T_{n}$:
\begin{equation}\label{change}
\begin{aligned}
& \Gamma=\frac{1}{N_{\text{reb}}}\sum_{n=1}^{N_{\text{reb}}} \sum_{i=1}^N |w^*_{i}(n) - w^*_{i}(n-1)|,
\end{aligned}
\end{equation}
where $N_{\text{reb}}$ is the number of rebalancing events (rebalancing here is performed every two months). 
The results are plotted in Figure \ref{trading}. As expected, $\Gamma$ is a monotonic function of $a$, which increases 10- to 20-fold between $a=0$ and the MVP case $a=1$. One interesting result is that the turnover is significantly reduced (by a factor $\sim 2$) when using a cleaned covariance matrix rather than its more volatile raw counterpart. 

\section{Agnostic Allocation: A Special Case of Particular Interest}

Out of the continuum of target portfolios proposed in the previous section, one choice of parameters plays a special role and leads to the so-called Agnostic Allocation Portfolio (AAP). In this section we recall the arguments of Benichou and al. \cite{ARP} that motivate such a choice. 

\subsection{Plain AAP}

AAP corresponds to $a=1/2$, $b=c=0$ in the construction of the target portfolio that the long-only version will be asked to track as closely as possible. As mentioned in section \ref{risk_decomp}, $a=1/2$ corresponds to the case where the risk is allocated proportionally to the natural projection of the predictor onto the eigenmodes. In order to justify this choice, Benichou et al. proposed to think about portfolio diversification in terms of symmetries rather than in terms of optimization \cite{ARP}. Let us consider returns $\bf{r}$ with a covariance matrix $\mathbf{C}$ and predictors ${\bf p}$ with a covariance matrix $\mathbf{Q}$. The idea is to create a set of independent, normalized synthetic assets and their corresponding predictors. Using a symmetry argument, the only rational allocation is then the following eigenrisk parity portfolio (ERP):
\begin{equation}\label{ARPtarget}
\begin{aligned}
w^{\text{t}}_{\text{ERP}} = \omega \mathbf{C}^{-1/2} \mathbf{Q}^{-1/2} \bf{p}
\end{aligned}
\end{equation}
This portfolio is such that the risk is equally spread on all the eigenmodes of the covariance matrix $\mathbf{C}$ \cite{ARP} -- hence the name ERP. 

Setting $\mathbf{Q} \equiv \mathbf{C}$ would only be warranted if one was certain that the predicted long-term excess returns ($\mathbf{p}$) behaved statistically similarly to the returns $\mathbf{r}$ themselves. In such a case, one recovers the standard Markowitz rule $w^{\text{t}}_{\text{ERP}}=w^{\text{t}}_{\text{MVP}}= \omega \mathbf{C}^{-1} \mathbf{p}$. However, since idiosyncratic excess returns are elusive and ephemeral, making assumptions about their correlations is treacherous. Benichou et al. \cite{ARP} suggested that the ``agnostic'' choice $\mathbf{Q} = \mathds{1}$ is a safer bet. 

For example, in the case of two correlated assets $A,B$ with correlation coefficient $\rho$ and a predictor that only lights up for asset $A$ and is zero for asset $B$, the Markowitz allocation puts a fraction $(1+\rho)/2$ of the risk on trading the spread $A-B$, betting that the difference in predictors will materialize in the future. Agnostic Allocation is more conservative and only allocates 50\% of the weight on the spread.   

This simple recipe has been shown to perform quite well when trading alternative beta strategies in futures, such as trend following or carry trades \cite{ARP}. In the present work, we advocate such a portfolio construction for long-only stock portfolios as well. In the following, we give a more detailed description of the AAP in the long-only case. We discuss the various merits of such a portfolio construction, and compare them with alternative portfolio constructions. As is already clear from Figures \ref{sparsity}-\ref{trading}, $a=1/2$ leads to a significant improvement in terms of portfolio concentration and turnover when compared to MVP.

\subsection{Eigen-Sparse AAP}

In this subsection, we want to motivate a further tweak to the AAP that limits its exposure to low-risk modes, for which the uniform predictor ${\bf p}=\mathds{1}$ has no intuitive interpretation. 

Indeed, let us study how ${\bf p}=\mathds{1}$ is decomposed over the eigenmodes of the covariance matrix. As expected, $\mathds{1}$ has a very large exposure to the top eigenvector of $\bf C$, $\textbf{u}_1$. This is often called the ``market mode'', with a vast majority of its components having the same sign over all stocks. Let us define the normalized component of ${\bf p}=\mathds{1}$ that is orthogonal to $\textbf{u}_1$ as the residual predictor $\mathds{1}_{\text{res}}$:
\begin{equation}\label{resid1}
\begin{aligned}
& \mathds{1}_{\text{res}}=\frac{\mathds{1}-(\mathds{1} \cdot \mathbf{u}_1) \mathbf{u}_1}{|\mathds{1}-(\mathds{1} \cdot \mathbf{u}_1) \mathbf{u}_1|}
\end{aligned}
\end{equation}
Its projection on the $k$th eigenmode of $\bf C$ is
\begin{equation}\label{projquad}
\begin{aligned}
& P_{\text{res}}(k) := (\mathds{1}_{\text{res}} \cdot \mathbf{u}_k)^2
\end{aligned}
\end{equation}
We want to study $P_{\text{res}}(k)$ as a function of $k \geq 2$. If $\mathds{1}_{\text{res}}$ and $\mathbf{u}_k$ were completely independent (i.e. the predictor $\mathds{1}$ is orthogonal to the $k$th mode), one would find 
$P_{\text{res},k} \sim 1/N$. In this case, it would make little sense to invest on mode $k$ based on predictor $\mathds{1}_{\text{res}}$. 

\begin{figure}
\begin{center}
\includegraphics[width=12cm]{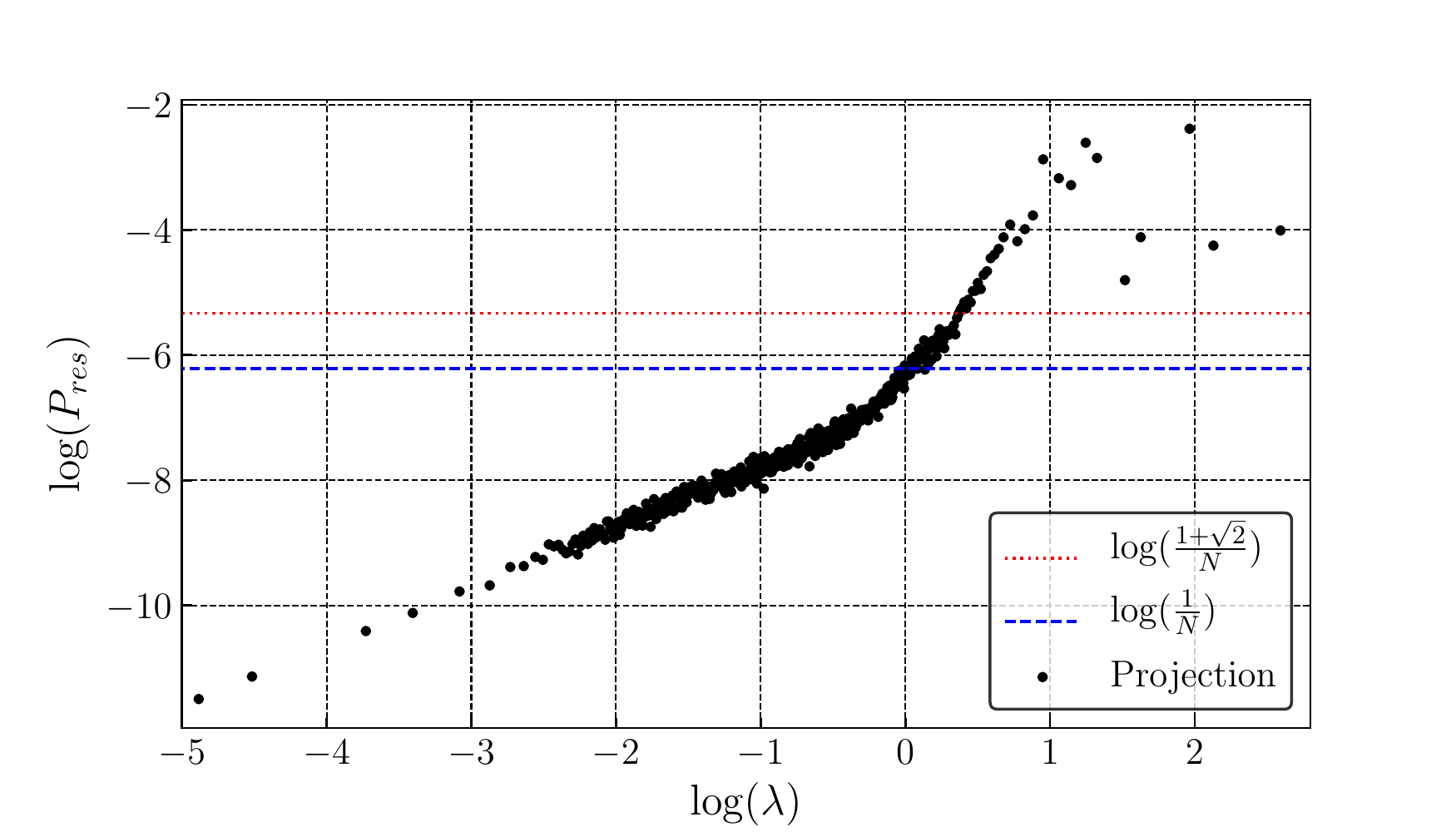}
\caption{Logarithm of the quadratic projection $P_{\text{res}}(k)$ of $\mathds{1}_{res}$ onto the eigenvectors of $\bf C$, as a function of the logarithm of the corresponding eigenvalues $\lambda_k$. $N=500$, $T=3500$, US stocks, 300 random selections. The two vertical lines correspond to $\log({1}/{N})$ and $\log(({1+\sqrt{2}})/{N})$. The value of $\lambda \approx 1.5$ where $P_{\text{res}}$ crosses the value ${(1+\sqrt{2})}/{N}$ corresponds to $k = k^* \approx 25$. For $k > k^*$, overlaps are deemed insignificant.}
\label{bootprojeig}
\end{center}
\end{figure}

The dependence of $P_{\text{res}}(k)$ is shown in Figure \ref{bootprojeig}. In order to obtain robust results, we have again used a bootstrap method by selecting 300 samples of $N=500$ US large cap stocks in the same pool of stocks already used. 

Figure \ref{bootprojeig} reveals the following interesting features (see also \cite{Allez}, section 6):
\begin{itemize}
\item $P_{\text{res}}(k)$ is of comparable magnitude for modes $k \lesssim
25$;
\item There is then a clear regular decrease of $P_{\text{res}}(k)$, which falls below the ``random'' threshold at the 1-sigma level $({1+\sqrt{2}})/{N}$ for $k \gtrsim 25$.
\item When conducted on other geographical zones, a similar behaviour holds with a threshold $k^*$ that corresponds to $\approx 5 \%$ of the size $N$ of the pool. 
\end{itemize}
These observations motivate us to truncate the AAP construction above $k^*$ in order to avoid exposure on modes with very little long-only predictability. Interestingly, this also has the effect of not trading large $k$, small risk modes that incur large impact costs.

So our final Eigen-Sparse Agnostic Allocation target portfolio reads:
\begin{equation}\label{cut}
\begin{aligned}
& w^{\text{t}}_{\text{S-AA}}=\omega\sum_{k=1}^{k^*} \frac{(\mathds{1} \cdot \mathbf{u}_k) \mathbf{u}_k}{\sqrt{\lambda_k}}
\end{aligned}
\end{equation}
with $k^* = 5\% \times N$. This target portfolio can be fed into the optimal tracking problem (\ref{optrb}) to get the Long-only AAP that we advocate.

\section{Long-Only AAP: Empirical Results}

\subsection{Methodology}

In this section we compare the performance and characteristics of the AAP to other risk-based methods across 5 major portfolios: US large caps, Europe (including the UK), Japan, Australia, and Canada. 

The construction of the trading pools is detailed in Appendix A. The competing methods are:

\begin{itemize}
    \item Agnostic Allocation Portfolio (AAP)
    \item (Eigen-) Sparse Agnostic Allocation Portfolio (S-AAP)
    \item Maximum Diversification Portfolio (MDP)
    \item Minimum Variance Portfolio (MVP)
    \item Equal Weight Portfolio (1/N)
    \item Market Cap Portfolio (MC)
\end{itemize}

Each method is re-balanced on a bi-monthly basis, and all share the same backward-looking and clean estimator of the covariance matrix (detailed in Appendix B). On top of the long-only constraint, we add a 3\% single position limit across all methods in order to ensure that the portfolio does not become overly concentrated on any single-name stock. Appendix C describes the optimization procedure in more details. We compute all performance metrics using weekly returns, but monthly returns also lead to very similar results. 

\subsection{Results and Interpretation}

We show in Table \ref{table:EUR} our results for European stocks, and other zones are reported in Appendix D. In terms of performance, we focus on:
\begin{enumerate}
    \item The annualized excess return (ER) over the risk-free rate
    \item The annualized total return (TR)
    \item The annualized volatility (Vol)
    \item The corresponding Sharpe ratio (SR)
\end{enumerate}
In terms of concentration and turnover, we report:
\begin{enumerate}
    \item The average number of non-zero positions 
    \item The average effective number of positions $N_{\text{eff}}$
    \item The annualized turnover, as defined below. 
    
\begin{equation}\label{LAT}
\begin{aligned}
& \text{Turnover}=\frac{1}{N_{\text{reb}}}\sum_{n=1}^{N_{\text{reb}}} \sum_{i=1}^N {|Z(n) w^*_{i}(n)-z_i(n) w^*_{i}(n-1))|^2},
\end{aligned}
\end{equation}
where $z_i(n)$ is the compounding factor for stock $i$ between rebalancing $n-1$ and $n$: $z_i(n):=\prod_{T_{n-1}<t\leq T_{n}}(1+r_i(t))$, while $Z:=\sum_i z_i/N$ is the average compounding factor. This definition is such that a market cap. based portfolio on a fixed pool of stocks has zero turnover.
\end{enumerate}

\begin{table}[!ht]
\centering
\begin{tabularx}{0.9\textwidth}{c@{\extracolsep\fill}r@{\extracolsep\fill}r@{\extracolsep\fill}r@{\extracolsep\fill}r@{\extracolsep\fill}r@{\extracolsep\fill}r}
\toprule
\multicolumn{7}{c}{EUR, Dec 1999 -- Dec 2018}   \\
\midrule
 & AAP & S-AAP & MDP & MVP & $\text{1/N}$ & MC \\
\midrule
ER & 6.4\% & 6.8\% & 4.8\% & 6.5\% & 5.7\% & 2.7\% \\
TR & 8.4\% & 8.8\% & 6.9\% & 8.5\% & 7.8\% & 4.7\% \\
Vol & 15.9\% & 16.8\% & 16.2\% & 12.6\% & 20.2\% & 19.8\% \\
SR & 0.40 & 0.40 & 0.30 & 0.51 & 0.28 & 0.13 \\
\midrule
No. Pos. & 831 & 861 & 88 & 79 & 868 & 860 \\
$N_{eff}$ & 338 & 544 & 50 & 45 & 868 & 158 \\
\midrule
Turnover & 146.4\% & 118.8\% & 235.2\% & 205.4\% & 74.8\% & 16.2\% \\

\midrule
$\rho$ & 93.7\% & 94.4\% & 82.9\% & 82.5\% & 96.3\% & 100.0\% \\
$\beta$ & 75.3\% & 80.1\% & 67.6\% & 52.2\% & 98.1\% & 100.0\% \\
$\alpha$ & 4.4\% & 4.6\% & 3.0\% & 5.1\% & 3.1\% & 0.0\% \\
\bottomrule
\end{tabularx}
\caption{Performance metrics for different portfolio constructions, for our European pool of stocks. Different columns correspond to different portfolios, rebalanced every 2 months, with an evolving pool of stocks based on liquidity. ER: excess return (over risk-free rate), TR: total return, Vol: volatility, SR: Sharpe Ratio. No. Pos.: average number of non-zero positions in the portfolio. Turnover: Average annual turnover, defined by Eq. (\ref{LAT}). $\rho, \beta, \alpha$: correlation, beta, and excess return with respect to the market cap. based portfolio (MC). Note that Turnover(MC) is positive as stocks enter and exit the pool.}
\label{table:EUR}
\end{table}

The salient features of our results are:
\begin{itemize}
    \item In terms of annualized returns and total returns, the best performers are S-AAP (Europe), $1/N$ and S-AAP (US) and MVP for smaller zones. Note however that $1/N$ and MVP have completely opposing properties in terms of volatility, concentration and turnover. 
    \begin{itemize}
        \item $1/N$ has the largest volatility, the largest effective number of assets (by definition) and the smallest turnover (except for the MC portfolio); 
        \item MVP, on the other hand, has the smallest volatility (by definition), but also the strongest concentration (low $N_{\text{eff}}$) and a large turnover. 
    \end{itemize}  
    \item Correspondingly, MVPs have, across the board, the highest risk-adjusted returns, or Sharpe ratios. But this comes at the expense of: 
        \begin{enumerate}
        \item High turnover, which is likely to strongly reduce returns after costs
        \item High concentration, that makes these portfolios susceptible to idiosyncratic ``black swans''. This is also the predicament of ``Maximally Diversified'' portfolios that are also maximally concentrated! 
        \item High exposure to the ``low-risk'' factor. This is already clear from the low average $\beta$ of MVP reported in Table \ref{table:EUR}, and further elaborated in Appendix E. 
    \end{enumerate} 
    \item $1/N$ portfolios, as well is known, perform surprisingly well out-of-sample \cite{1N}. They also trade very little, but their volatility is so high that their Sharpe ratio is always smaller than our AAP alternative.
    \item AAP and even more so Sparse-AAP display an interesting compromise between all desired features, i.e. in terms of excess and total returns, concentration, and turnover. For all zones except Australia (which is the smallest pool), the volatility of S-AAP is smaller than that of the market capitalization (MC) portfolio, and the total returns are larger, all the while maintaining a reasonable level of turnover, concentration and exposure to low-risk factors. 
\end{itemize}

\section{Conclusion}

Agnostic Allocation Portfolios represent a good compromise between two broad families of risk-based portfolios: those structurally allocating to the full investment universe and those operating some stock selection while seeking to minimize a risk-related quantity, often relying on the inverse of the covariance matrix of returns.

We have argued theoretically in favor of the (eigen-) Sparse Agnostic Allocation Portfolio construction, and established empirically that long-only S-AAP, with no further signals, offers similar to better risk-adjusted performance than standard alternatives such as MDP or MVP, especially for large pools. Additionally, the S-AAP is much less concentrated than its optimization-based competitors, and thus it is less exposed to idiosyncratic risk. Finally, S-AAP is much less demanding in terms of portfolio turnover and transaction costs (while still trading $\sim 100 \%$ of its assets on a yearly basis). We also focused on implementation efficiency: concentration effects and excess trading can be substantially reduced by using adequately cleaned covariance matrices.  

We believe that our general Agnostic Allocation framework (originally devised in \cite{ARP}) is relevant in many other situations. For example, we have not considered in this paper adding active quantitative equity strategies (for example, Momentum or Value). One could expect AAPs to outperform their Markowitz counterparts (see \cite{ARP} for a discussion in the case of trend following in the futures space). Another interesting application would be to create a long-only Agnostic Allocation strategy using the whole universe of futures contracts. We leave these extensions for future work. 

\subsection*{Acknowledgments} We thank G. Azevedo, R. Benichou, A. Beveratos, L. De Leo, S. Gualdi, C. Lehalle, M. Potters, G. Simon for help, inspiration and very insightful comments.





\section*{Appendix A: pools of instruments}

Results are computed for the geographical zones below. For each geography, we update the pool of instruments on a yearly basis based on a backward looking 3 month liquidity filter. At each rebalancing date we only consider stocks with 95\% of available returns over the look back period used to compute the covariance matrix (1000 days).  

\begin{itemize}
\item US: we consider the 1000 most liquid stocks in the Russell 3000 Index. 
\item Canada: we  consider the top 500 largest cap stocks from which we extract the 200 most liquid stocks.
\item Europe: this zone includes the UK in addition to developed markets in continental
Europe. We consider the 2000 largest cap stocks and then select the 1000 most liquid stocks among them.
\item Japan: we select the 1000 most liquid stocks in the TOPIX index.
\item Australia: we take the 500 largest cap stocks and then select the 200 most liquid stocks among them.

\end{itemize}

\section*{Appendix B: cross-validated eigenvalues \& algorithm}
\label{appendix:B}
 Covariance matrices are key for objective-based portfolio construction. Since the ``true" covariance matrix is never known, and in the absence of good priors,  one has to rely on that measured using past data. 
 
If $r_i(t)$ is the daily return of stock $i$ at time $t$, the empirical covariance for stocks $i$ and $j$ over time $T$ (neglecting daily mean returns) is written as:
\begin{equation}\label{Covar}
\begin{aligned}
& C_{i,j}^{\text E}=\frac{1}{T} \sum_{t=1}^{\text{t}} r_i(t) r_j(t),
\end{aligned}
\end{equation}
that we write in matrix form as
\begin{equation}\label{Covarmat}
\begin{aligned}
& \operatorname{C}^{\text E}=\frac{1}{T} \operatorname{R}\operatorname{R}^\top 
\end{aligned}
\end{equation}
with $R$ being the $N$ by $T$ rectangle matrix of returns.
In the diagonal basis of eigenvectors, one has
\begin{equation}\label{Covarmatdiag}
\begin{aligned}
& \operatorname{C}^{\text E} = 
\sum_{k=1}^N \lambda_k^{\text E} u_k^{{\text E}\top} u_k^{{\text E}}
\end{aligned}
\end{equation}
where
$\lambda_1^{\text E} \leq \lambda_2^{\text E}\leq...\leq\lambda_N^{\text E}\leq0$ are the eigenvalues associated with eigenvectors $u_1^{\text E},u_2^{\text E},...,u_N^{\text E}$ of $C^{\text E}$. 

When ${q}= N/T \to 0$, i.e., when the return data covers a very long history, the empirical covariance matrix is expected to converge toward the true covariance matrix (if such a thing exists!). For realistic samples (ie $N \sim T$), the finite nature of the time series introduces noise and systematic errors in the computation of eigenvalues. The smallest eigenvalues are systematically underestimated, leading to sub-optimal portfolios, spuriously over-allocated to low risk configurations. These also incur more concentration and turnover. 

The benefits of cleaning correlation matrices are clearly established and multiple methods are available: for a review, see e.g. \cite{cleancor} and references therein. For the purpose of this paper we use a cross-validation based estimation of eigenvalues \cite{Bun}. It is particularly adaptable as it allows us to cover cases where $T<N$ and ensures strictly positive values for eigenvalues regardless of potential missing data issues. Below is a pseudo-code algorithm for computing cross-validated eigenvalues. 

Let $\mathbf{R}$ be a matrix of standardized returns across a period of $T$ for $N$ instruments. The \textit{non-normalized} covariance matrix over the full period is
\begin{align}
    \widehat{C}_{ij} = \sum_{t}^{T} r_i(t) r_j(t) =  (\mathbf{R}\mathbf{R}^\top)_{ij}
\end{align}

The algorithm computes at each time step an independent estimation of the variance on a withheld portion of the time series of returns. At each iteration, we randomly sample without replacement $10\%$ of days $(T_\text{out})$ to construct a new return matrix $\mathbf{R}^\text{out} \in \mathbb{R}^{N \times T_\text{out}}$. We can then compute the \textit{non-normalized} covariance matrix on the withheld days.
\begin{align}
    \widehat{C}_{ij}^\text{out} = \sum_{\tau}^{T_{\text{out}}} r_{i}(\tau) r_{j}(\tau) =  (\mathbf{R}^{\text{out}} {\mathbf{R}^{\text{out}}}^\top)_{ij}
\end{align}
The covariance matrix on the ``main" (non withheld) period is
\begin{align}
    \widehat{C}_{ij}^\text{in} &= \frac{1}{T - T_{\text{out}}} \bigg(\sum_{t}^{T} r_i^{\text{t}} r_j^{\text{t}} - \sum_{\tau}^{T_{\text{out}}} r_{i}^\tau r_{j}^\tau \bigg)\\ 
    &= \frac{1}{T - T_{\text{out}}} \bigg( (\mathbf{R}\mathbf{R}^\top)_{ij} - (\mathbf{R}^{\text{out}} \mathbf{R}^{\text{out}^\top})_{ij} \bigg)
\end{align}

We compute the eigenvectors $\mathbf{\widehat{u}}_k^{\text{in}}$ of $ \widehat{\mathbf{C}}^{\text{in}} $ and then compute their variance  on the ``left out" returns, thus providing an ``out-of-sample" estimation of the eigenvalues for $\mathbf{\widehat{u}}_k^{\text{in}}$:
\begin{align}
    \widehat{\lambda}_k := \frac{1}{T_\text{out}}\sum_{\tau}^{T_{\text{out}}} \sum_{k}^N (\mathbf{\widehat{u}}_k^{\text{in}} {r}_k^{\tau})^2 
\end{align}

\begin{figure}[!ht]
\begin{center}
\includegraphics[width=11cm]{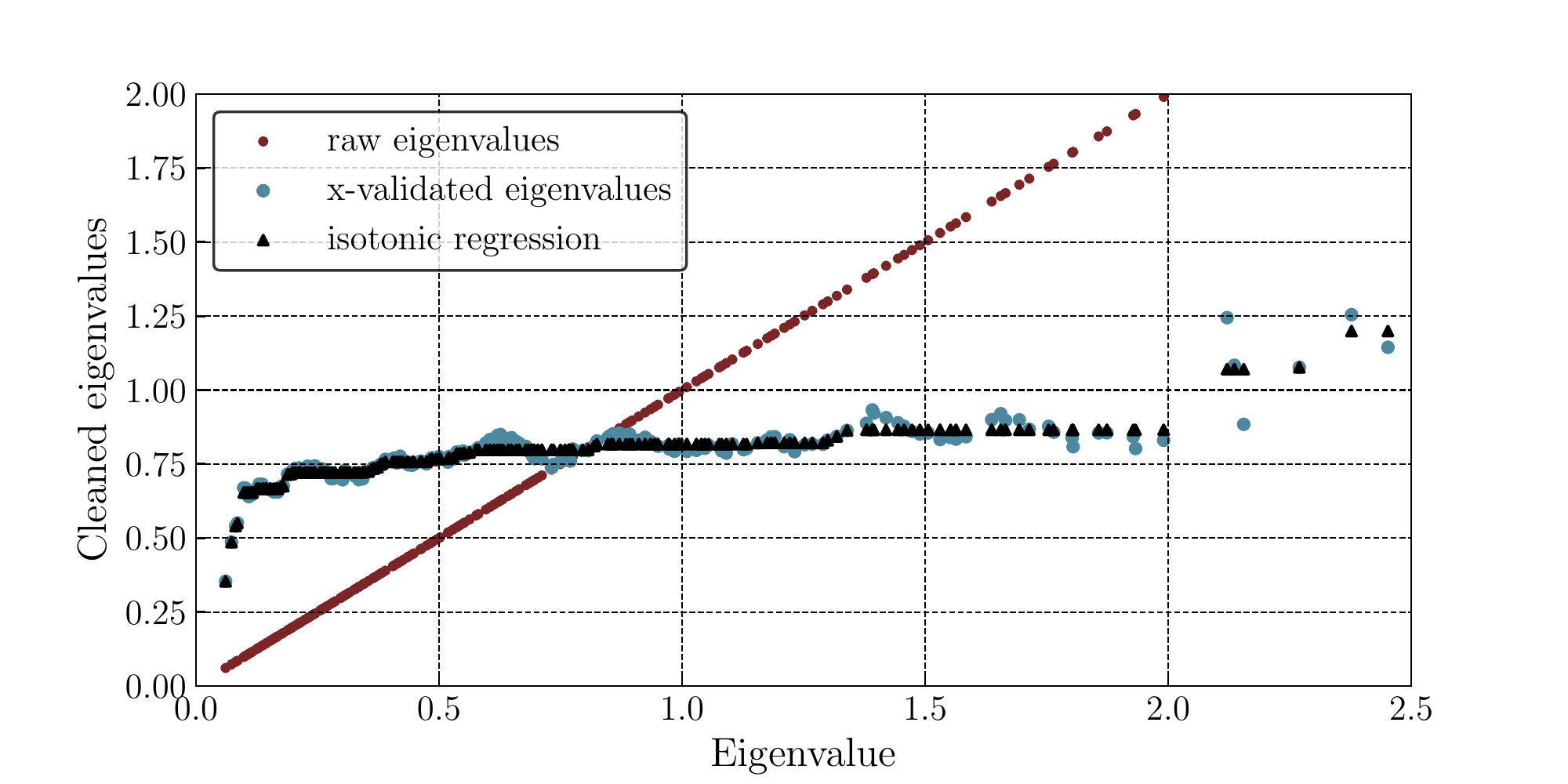}
\caption{``Cleaned'' eigenvalues as a function of ``in-sample'' eigenvalues. Red: no cleaning. Blue: Cross validated. Black: Cross validated using an isotonic smoothing. This is a zoom on low-risk eigenvalues enabling us to see the correction of small eigenvalues (underestimated), and larger eigenvalues (over estimated).}
\label{cv}
\end{center}
\end{figure}

This process is repeated $T$ times. The resulting out-of-sample eigenvalues are averaged over $T$. We make an isotonic fit of the cross-validated eigenvalues as a function of the in-sample ones, in order to impose the hierarchy of the eigenvalues measured on the full sample \cite{Marcinprep}. Figure \ref{cv} illustrates the results of such a procedure. The Python code will be made available in a notebook upon request. 

\section*{Appendix C: Portfolio Construction Protocol for AAP and Objective function based portfolios}
\subsection*{Returns}
We rely on the universe of assets described in Appendix A, and on CFM's database of global stock returns which is carefully assembled for production purposes. The covariance matrices computed for the AAP and other optimization-based methods use total returns $r_i(t)$ (i.e. adjusted for dividends and other corporate actions), which are cross-sectionally normalized into $\widetilde{r}_i(t)$. This is a typical adjustment attempting to ``stationnarize'' the return time series and avoid overweighting high volatility periods in the sample. More precisely; 

\begin{equation}\label{CSrenorm}
\widetilde{r}_i(t):=\frac{r_i(t)}{\sqrt{\sum_{i=1}^{N}(r_i(t))^2}} 
\end{equation}
For all zones, we use returns in US Dollars, hence placing ourselves in the shoes of a US investor. 

\subsection*{Rebalancing and optimization}

For comparative purposes, we run all optimization methods every other month-end for each geographical zone. Results would not materially be affected by a monthly (or daily) update, however more frequent updates result in higher annual turnover (and thus costs). 

At each optimization date we compute covariance matrices for the instruments in the trading pool over the 1000 previous trading days (4 years). A long look-back period is favored as it naturally reduces turnover. We introduce a safe 2-day lag making sure that we use information that is available on the optimization date. Stocks that are included in the trading pool but having more than 50 (5\%) unavailable prices are excluded from the portfolio. 

In order to put all methods on an equal footing, we compute covariance matrices using the cross-validation method described in Appendix B for all of them. As discussed in the body of the paper, this has a significant effect on the turnover of the portfolio, in particular for ``$a=1$" methods (MVP and MDP). 

The optimization problem that is solved at every optimization date is (\ref{optrb}) for the relevant target portfolios $\bf{w}^{\text{t}}$, with a tighter single-name constraint at 3\%: this is a reasonable level for a realistic implementation of a risk-based portfolio. This constraint will mostly have an effect on the ``$a=1$'' methods, by helping them avoid extremely concentrated configurations. Constraints corresponding to concentration limits in various jurisdictions (e.g. UCITS) can easily be added. 

\begin{equation}\label{optrbfinal}
\begin{aligned}
 {\bf w}^*= & \enskip \operatorname*{arg\,min}_w   & &({\bf w}-{\bf w}^{\text{t}})^\top \mathbf{C} ({\bf w}-{\bf w}^{\text{t}})\\
& \text{subject to} & & \enskip 0 \leq  w_i \leq 0.03\sum_{j=1}^{N}w_j
\end{aligned}
\end{equation}
The Python code and documentation for this optimization problem will also be made available in a notebook upon request.

\newpage

\section*{Appendix D: Tables for US, JP, CAN \& AUS}
Captions: see Table \ref{table:EUR}
\begin{table}[!ht]
\centering
\begin{tabularx}{0.9\textwidth}{c@{\extracolsep\fill}r@{\extracolsep\fill}r@{\extracolsep\fill}r@{\extracolsep\fill}r@{\extracolsep\fill}r@{\extracolsep\fill}r}
\toprule
\multicolumn{7}{c}{US, Dec 1993 -- Dec 2018}   \\
\midrule
 & AAP & S-AAP & MDP & MVP & $\text{1/N}$ & MC \\
\midrule
ER & 7.1\% & 7.3\% & 5.0\% & 6.7\% & 7.5\% & 6.0\% \\
TR & 10.1\% & 10.3\% & 7.9\% & 9.6\% & 10.5\% & 8.9\% \\
Vol & 14.3\% & 16.6\% & 15.4\% & 11.2\% & 19.6\% & 17.1\% \\
SR & 0.5 & 0.44 & 0.32 & 0.6 & 0.38 & 0.35 \\
\midrule
No. Pos. & 765 & 812 & 117 & 82 & 819 & 816 \\
$N_{\text{eff}}$ & 279 & 500 & 56 & 44 & 819 & 167 \\
\midrule
Turnover & 143.1\% & 139.9\% & 231.8\% & 172.7\% & 92.8\% & 19.9\% \\

\midrule
$\rho$ & 93.4\% & 94.6\% & 80.5\% & 76.9\% & 95.6\% & 100.0\% \\
$\beta$ & 77.9\% & 91.7\% & 72.7\% & 50.1\% & 109.7\% & 100.0\% \\
$\alpha$ & 2.4\% & 1.8\% & 0.7\% & 3.7\% & 0.9\% & 0.0\% \\
\bottomrule
\end{tabularx}
\end{table}

\begin{table}[!ht]
\centering
\begin{tabularx}{0.9\textwidth}{c@{\extracolsep\fill}r@{\extracolsep\fill}r@{\extracolsep\fill}r@{\extracolsep\fill}r@{\extracolsep\fill}r@{\extracolsep\fill}r}
\toprule
\multicolumn{7}{c}{JP, Sep 1997 -- Jan 2019}   \\
\midrule
 & AAP & S-AAP & MDP & MVP & $\text{1/N}$ & MC \\
\midrule
ER & 2.0\% & 1.9\% & -1.6\% & 2.8\% & 1.8\% & 0.0\% \\
TR & 4.4\% & 4.3\% & 0.7\% & 5.2\% & 4.2\% & 2.3\% \\
Vol & 18.2\% & 18.8\% & 20.5\% & 15.4\% & 21.0\% & 19.8\% \\
SR & 0.11 & 0.1 & -0.08 & 0.18 & 0.09 & 0.0 \\
\midrule
No. Pos. & 472 & 474 & 62 & 53 & 474 & 474 \\
$N_{\text{eff}}$ & 232 & 346 & 44 & 39 & 474 & 100 \\
\midrule
Turnover & 133.2\% & 119.9\% & 213.5\% & 170.1\% & 85.7\% & 12.4\% \\

\midrule
$\rho$ & 94.0\% & 94.8\% & 85.0\% & 79.3\% & 95.4\% & 100.0\% \\
$\beta$ & 86.3\% & 90.0\% & 87.6\% & 61.7\% & 100.7\% & 100.0\% \\
$\alpha$ & 2.0\% & 2.0\% & -1.6\% & 2.8\% & 1.8\% & 0.0\% \\
\bottomrule
\end{tabularx}
\end{table}

\begin{table}[!ht]
\centering
\begin{tabularx}{0.9\textwidth}{c@{\extracolsep\fill}r@{\extracolsep\fill}r@{\extracolsep\fill}r@{\extracolsep\fill}r@{\extracolsep\fill}r@{\extracolsep\fill}r}
\toprule
\multicolumn{7}{c}{CAN, Dec 1999 -- Dec 2018}   \\
\midrule
 & AAP & S-AAP & MDP & MVP & $\text{1/N}$ & MC \\
\midrule
ER & 8.4\% & 7.7\% & 7.0\% & 10.3\% & 6.6\% & 7.0\% \\
TR & 10.5\% & 9.8\% & 9.1\% & 12.5\% & 8.7\% & 9.1\% \\
Vol & 17.7\% & 20.0\% & 19.7\% & 14.9\% & 22.2\% & 20.3\% \\
SR & 0.47 & 0.38 & 0.36 & 0.69 & 0.3 & 0.34 \\

\midrule
No. Pos. & 164 & 164 & 62 & 49 & 164 & 161 \\
$N_{\text{eff}}$ & 97 & 133 & 43 & 38 & 164 & 47 \\
\midrule
Turnover & 115.4\% & 105.4\% & 193.8\% & 134.9\% & 92.7\% & 16.7\% \\

\midrule
$\rho$ & 95.3\% & 96.4\% & 89.4\% & 89.7\% & 96.0\% & 100.0\% \\
$\beta$ & 83.0\% & 94.9\% & 86.7\% & 65.8\% & 104.8\% & 100.0\% \\
$\alpha$ & 2.6\% & 1.0\% & 1.0\% & 5.7\% & -0.7\% & 0.0\% \\
\bottomrule
\end{tabularx}
\end{table}

\begin{table}[!ht]
\centering
\begin{tabularx}{0.9\textwidth}{c@{\extracolsep\fill}r@{\extracolsep\fill}r@{\extracolsep\fill}r@{\extracolsep\fill}r@{\extracolsep\fill}r@{\extracolsep\fill}r}
\toprule
\multicolumn{7}{c}{AUS, Dec 1999 -- Dec 2018}   \\
\midrule
 & AAP & S-AAP & MDP & MVP & $\text{1/N}$ & MC \\
\midrule
ER & 8.6\% & 8.1\% & 6.5\% & 10.0\% & 7.4\% & 9.2\% \\
TR & 10.7\% & 10.2\% & 8.6\% & 12.1\% & 9.5\% & 11.3\% \\
Vol & 22.0\% & 23.4\% & 25.9\% & 19.4\% & 23.7\% & 22.5\% \\
SR & 0.39 & 0.34 & 0.25 & 0.52 & 0.31 & 0.41 \\
\midrule
No. Pos. & 157 & 157 & 49 & 50 & 157 & 154 \\
$N_{\text{eff}}$ & 109 & 143 & 39 & 39 & 157 & 27 \\
\midrule
Turnover & 123.7\% & 102.2\% & 211.7\% & 170.1\% & 87.3\% & 19.1\% \\

\midrule
$\rho$ & 95.1\% & 96.0\% & 88.6\% & 92.6\% & 96.3\% & 100.0\% \\
$\beta$ & 92.8\% & 99.8\% & 101.8\% & 79.7\% & 101.3\% & 100.0\% \\
$\alpha$ & 0.1\% & -1.1\% & -2.8\% & 2.7\% & -1.9\% & 0.0\% \\
\bottomrule
\end{tabularx}
\end{table}

\newpage{}

\section*{Appendix E: Exposure to the low-vol and low-beta factors}

We provide an analysis of the low-risk biases of the RBPs introduced in this article, aiming at accounting for the outperformance of MPV in terms of risk-adjusted returns. The low-risk anomaly is documented as a robust and persistent factor of out-performance \cite{Ang, Ang2}, although mostly explained by the dividend-based value factor (see e.g. \cite{lowvol}). 

\subsection*{Definition}

We measure the exposure of the tested portfolios to two well-documented ``low risk" factor: low-vol and low-beta. In order to construct times series of returns for those factors, we essentially follow the procedure described in \cite{lowvol}. Volatility ($\sigma_i$) and the beta ($\beta_i$) of each stock is measured as a 100-day rolling standard deviation of the stocks returns. Betas are measured as the covariance of each stock to the MC portfolio in each zone (the index), divided by the index variance, both computed over 100 days. We consider 3-days returns for the variances and covariances to account for any lead-lag effects, and we lag these values by 20 trading days to exclude short-term reversals. Volatility and Beta are then ranked and shifted in order to ensure cash-neutrality for the signal: 

\begin{equation}\label{rank}
\begin{aligned}
 s_i(t)=  \frac{2}{N}\textrm{rank}\bigg(\frac{1}{\sigma_i(t)}\bigg)-1
\end{aligned}
\end{equation}

We then construct the corresponding time series of returns. To compensate the structural short market bias of the low risk portfolios we compute the 100-day beta of those time series to the corresponding MC portfolios, with a 2-day lag in order to ensure causality. Additionally, in order to make the P\&L stationary, we control its volatility to a 10\%  (arbitrary) target using a lagged 100-day rolling volatility estimate.  

\subsection*{Exposures and interpretation}

We measure for each the low-vol and low-beta factor their correlation to: 
\begin{itemize}
    \item the returns of all competing risk based portfolios ($\rho$). 
    \item the returns of all competing portfolios adjusted for their (2-day lagged rolling 100-day) beta to the MC portfolios ($\rho^*$), thus representing the ``idiosyncratic" excess return of each portfolio. 
\end{itemize}

We summarize below the main results: 

\begin{itemize}
    \item Across all geographical regions, both low risk factors are highly correlated to the MVP. This is indicative of a consistent low risk bias of this portfolio construction method. The exposure to low-beta is higher than that of low-vol.  
    \item MDP, on the other hand, has a robust {\it negative} exposure to low-risk, particularly acute in the low-vol factor. It is an expected feature of that method which emphasizes high volatility stocks relative to MVP. 
    \item The residual of the 1/N portfolio has a strong negative correlation to both low risk factors, thus reflecting its higher small-cap exposure relative to the benchmark. 
    \item S-AAP and AAP both exhibit, overall, a more neutral exposure to low-risk factors.
\end{itemize}

\begin{table}[!ht]
\centering
\begin{tabularx}{0.75\textwidth}{c@{\hskip 0.2in}l@{\extracolsep\fill}r@{\extracolsep\fill}r@{\extracolsep\fill}r@{\extracolsep\fill}r@{\extracolsep\fill}r}
\toprule
  &  & AAP & S-AAP & MDP & MVP & $\text{1/N}$  \\
\midrule
\multirow{4}{*}{EUR} & $\rho_{LV}$ & 4.2\% & 3.8\% & -0.8\% & 20.1\% & -3.2\% \\
& $\rho^*_{LV}$ & -7.9\% & -9.5\% & -12.0\% & 26.1\% & -42.0\%\\
\cmidrule(){2-7}
& $\rho_{L\beta}$ & 8.9\% & 7.6\% & 9.8\% & 26.6\% & -2.1\% \\ 
& $\rho^*_{L\beta}$ & 20.3\% & 17.4\% & 14.5\% & 45.5\% & -19.7\%\\ 
\midrule
\multirow{4}{*}{US} & $\rho_{LV}$ & 5.1\% & 3.0\% & -7.5\% & 24.8\% & -5.2\% \\
& $\rho^*_{LV}$ & -14.6\% & -23.7\% & -28.9\% & 31.7\% & -61.0\% \\
\cmidrule(){2-7}
& $\rho_{L\beta}$ &  7.3\% & 4.8\% & 0.6\% & 26.0\% & -4.8\% \\ 
& $\rho^*_{L\beta}$& 2.3\% & -7.1\% & -10.0\% & 38.5\% & -48.7\% \\ 
\midrule
\multirow{4}{*}{JP} & $\rho_{LV}$ &   12.1\% & 11.4\% & 2.2\% & 31.8\% & 5.3\%\\
& $\rho^*_{LV}$ &  -6.0\% & -9.6\% & -22.2\% & 40.2\% & -38.6\% \\
\cmidrule(){2-7}
& $\rho_{L\beta}$ & 19.2\% & 17.8\% & 16.8\% & 40.2\% & 9.1\% \\ 
& $\rho^*_{L\beta}$& 22.2\% & 17.7\% & 9.7\% & 57.9\% & -19.7\%\\ 
\midrule
\multirow{4}{*}{CAN} & $\rho_{LV}$ & -0.4\% & -4.3\% & -14.0\% & 17.0\% & -13.4\% \\
& $\rho^*_{LV}$ & -13.2\% & -29.5\% & -38.9\% & 32.5\% & -64.0\% \\
\cmidrule(){2-7}
& $\rho_{L\beta}$ &  4.1\% & -0.5\% & -4.1\% & 19.8\% & -9.7\%  \\
& $\rho^*_{L\beta}$& 2.9\% & -14.0\% & -15.9\% & 39.6\% & -49.3\%\\ 
\midrule
\multirow{4}{*}{AUS} & $\rho_{LV}$ & 0.6\% & -0.1\% & -18.0\% & 13.8\% & -2.5\%\\
& $\rho^*_{LV}$ & -31.7\% & -39.4\% & -57.5\% & 10.5\% & -49.6\%\\
\cmidrule(){2-7}
& $\rho_{L\beta}$ & 5.2\% & 2.6\% & -5.9\% & 17.0\% & 0.0\%\\ 
& $\rho^*_{L\beta}$& -4.6\% & -15.6\% & -25.0\% & 28.3\% & -26.2\%\\ 

\bottomrule
\end{tabularx}

\caption{Correlation of the low-vol (LV) and low-beta (L$\beta$) factors with each portfolio construction method ($\rho$), and with their residual over the MC portfolio ($\rho*$).}
\label{table:LOW}
\end{table}

\newpage
\bibliographystyle{abbrv}

\begin{thebibliography}{}

\end{thebibliography}


\begin{thebibliography}{99}
\bibitem{ARP} Benichou, R., Lemperiere, Y., Sérié,E., Kockelkoren, J., Seager, P., Bouchaud, J.P. \& Potters, M.
\href{ }{\emph{ Agnostic Risk Parity: Taming Known and Unknown-Unknowns.}} The Journal of Investment Strategies. 6. 10.21314,
  (2016)

\bibitem{MV0} Haugen, R. \& Baker, N. 
\href{ }{\emph{ The Efficient Market Inefficiency of Capitalization-Weighted Stock Portfolios.}} Journal of Portfolio Management. 17(3), 35–40,
  (1991)

\bibitem{MV1} Clarke, R., de Silva, H. \& Thorley, S.
\href{ }{\emph{ Minimum-Variance Portfolios in the U.S. Equity Market.}} Journal of Portfolio Management. 33. 10-24.,
 (2006)
 
 \bibitem{MV2} Clarke, R., De Silva, H. \& Thorley, S.  \href{ }{\emph{ Minimum Variance Portfolio
Composition.}} Journal of Portfolio Management, 31(2):31-45 (2011)

\bibitem{ERC} Maillard, S., Roncalli, R. \& Teiletche, J.
\href{ }{\emph{The Properties of Equally Weighted Risk
Contribution Portfolios.}} Journal of Portfolio Management, 36, 60-70., (2010)

\bibitem{RBI}  Demey, P., Maillard, S. \& Roncalli, T. \href{ }{\emph{ Risk-Based Indexation. Lyxor White Paper.}} https://ssrn.com/abstract=1582998, (2010)

\bibitem{MDP}  Choueifaty, Y. \&  Coignard, Y.
\href{ }{\emph{ Toward Maximum Diversification.}} Journal of Portfolio Management. 35. 40-51.,
  (2008)

\bibitem{1N}  Windcliff, H. \& Boyle P., 
\href{ }{\emph{ The 1/n pension plan puzzle,}} North American Actuarial Journal. 8. 32-45,
(2004)

\bibitem{Michaud} Michaud, R. O., \href{ }{\emph{ The Markowitz optimization enigma: Is ‘optimized’ optimal?.}} Financial Analysts Journal, 45(1), 31-42. (1989)


\bibitem{CAL}  Lehalle, C.A. \& Simon G., 
\href{ }{\emph{ Portfolio Selection With Active Strategies: How Long Only Constraints Shape Convictions.}} Working paper,
(2019)

\bibitem{cleancor}  Bun, J., Bouchaud, J.P. \& Potters, M.,
\href{ }{\emph{Cleaning correlation matrices,}} Risk magazine,
(2016)

\bibitem{Fernandez} Fernandez, P. \href{ }{\emph{ CAPM: an absurd model.}} Business Valuation Review, 2015, vol. 34, no 1, p. 4-23.

\bibitem{Allez} Allez, R., \& Bouchaud, J. P., \href{ }{\emph{ Eigenvector dynamics: general theory and some applications.}} Physical Review E, 86(4), 046202  (2012)

\bibitem{Bun} Bun, J., Bouchaud, J. P., \& Potters, M.,\href{ }{\emph{ Overlaps between eigenvectors of correlated random matrices.}} Physical Review E, 98(5), 052145 (2018)

\bibitem{Marcinprep} M. Potters et al., in preparation (2019).

\bibitem{Ang} Ang, A., Hodrick, R. J., Xing, Y., \& Zhang, X. \href{ }{\emph{ The cross‐section of volatility and expected returns.}} The Journal of Finance, 61(1), 259-299 (2006).

\bibitem{Ang2} Ang, A., Hodrick, R. J., Xing, Y., \& Zhang, X. \href{ }{\emph{ High idiosyncratic volatility and low returns: International and further US evidence.}} Journal of Financial Economics, 91(1), 1-23 (2009).


\bibitem{lowvol} Beveratos, A., Bouchaud, J. P., Ciliberti, Laloux, L., Lempérière, Y., Potters, M., \& Simon, G.,\href{ }{\emph{ Deconstructing the Low-Vol Anomaly.}} The Journal of Portfolio Management, 44 (1) 91-103 (2017)




\end{thebibliography}

\end{document}